# 6G Wireless Communication Systems: Applications, Requirements, Technologies, Challenges, and Research Directions

Mostafa Zaman Chowdhury[1,2], Md. Shahjalal[1], Shakil Ahmed[3], and Yeong Min Jang[1]
[1]Dept. of Electronics Engineering, Kookmin University, Seoul, Republic of Korea
[2]Dept. of Electrical and Electronic Engineering, Khulna University of Engineering & Technology, Khulna, Bangladesh
[3]Dept. of Electrical and Computer Engineering, The University of Arizona, USA
E-mail: mzaman@kookmin.ac.kr, {mdshahjalal26, shakilahmed}@ieee.org, yjang@kookmin.ac.kr

*Abstract*—Fifth-generation (5G) communication, which has many more features than fourth-generation communication, will be officially launched very soon. A new paradigm of wireless communication, the sixth-generation (6G) system, with the full support of artificial intelligence is expected to be deployed between 2027 and 2030. In beyond 5G, there are some fundamental issues, which need to be addressed are higher system capacity, higher data rate, lower latency, and improved quality of service (QoS) compared to 5G system. This paper presents the vision of future 6G wireless communication and its network architecture. We discuss the emerging technologies such as artificial intelligence, terahertz communications, optical wireless technology, free space optic network, blockchain, three-dimensional networking, quantum communications, unmanned aerial vehicle, cell-free communications, integration of wireless information and energy transfer, integration of sensing and communication, integration of access-backhaul networks, dynamic network slicing, holographic beamforming, and big data analytics that can assist the 6G architecture development in guaranteeing the QoS. We present the expected applications with the requirements and the possible technologies for 6G communication. We also outline the possible challenges and research directions to reach this goal.

## I. Introduction

We are moving toward a society of fully automated and remote management systems. The very rapid development of various emerging applications, such as artificial intelligence (AI), virtual reality, three-dimensional (3D) media, and the internet of everything (IoE), has led to a massive volume of traffic [1]. The global mobile traffic volume was 7.462 EB/month in 2010, and this traffic is predicted to be 5016 EB/month in 2030 [2]. This statistic clearly depicts the importance of the improvement of communication systems. Autonomous systems are becoming popular in every sector of society, such as industry, health, roads, oceans, and space. To provide a smart life and automated systems, millions of sensors will be embedded into cities, vehicles, homes, industries, foods, toys, and other environments. Hence, a high-data-rate with reliable connectivity will be required to support these applications. In certain parts of the world, fifth-generation (5G) wireless networks have already been deployed. By 2020, it is expected that 5G will be fully deployed worldwide.

5G networks will not have the capacity to deliver a completely automated and intelligent network that provides everything as a service and a completely immersive experience [3]. Although, the 5G communication systems that are going to be released very soon will offer significant improvements over the existing systems, they will not be able to fulfill the demands of future emerging intelligent and automation systems after 10 years [4]. The 5G network will provide new features and provide better quality of service (QoS) as compared with fourth-generation (4G) communications [5]–[8]. The 5G technology will include several new additional techniques, such as new frequency bands (e.g., the millimeter wave (mmWave) and the optical spectra), advanced spectrum usage and management, and the integration of licensed and unlicensed bands [4]. Nevertheless, the fast growth of data-centric and automated systems may exceed the capabilities of 5G wireless systems. Certain devices, such as virtual reality (VR) devices need to go beyond 5G (B5G) because they require a minimum of 10 Gbps data rate [1]. Hence, with 5G reaching its limits in 2030, the design goals for its next step are already being explored in literature.

To overcome the constraints of 5G for supporting new challenges, a sixth-generation (6G) wireless system will need to be developed with new attractive features. The key drivers of 6G will be the convergence of all the past features, such as network densification, high throughput, high reliability, low energy consumption, and massive connectivity. The 6G system would also continue the trends of the previous generations, which included new services with the addition of new technologies. The new services include AI, smart wearables, implants, autonomous vehicles, computing reality devices, sensing, and 3D mapping [9]. The most important requirement for 6G wireless networks is the capability of handling massive volumes of data and very high-data-rate connectivity per device [1].

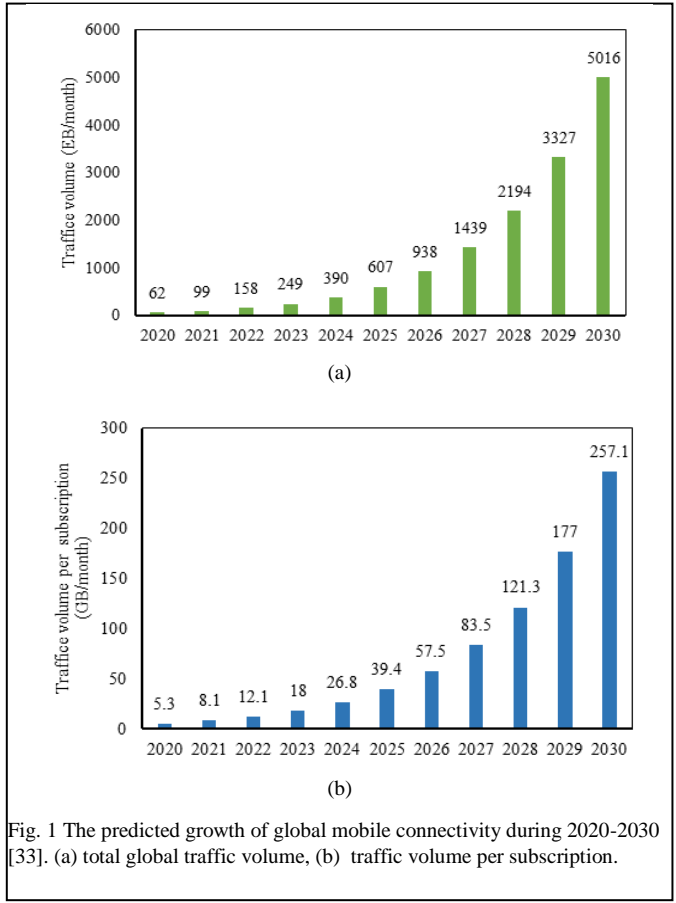

Fig. 1 The predicted growth of global mobile connectivity during 2020-2030 [33]. (a) total global traffic volume, (b) traffic volume per subscription.

The 6G system will increase performance and maximize user QoS several folds more than 5G along with some exciting features. It will protect the system and secure the user data. It will provide comfortable services [10]. The 6G communication system is expected to be a global communication facility. It is envisioned that the per-user bit rate in 6G will be approximately 1 Tb/s in many cases [1], [11]. The 6G system is expected to provide simultaneous wireless connectivity that is 1000 times higher than 5G. Moreover, ultra-long-range communication with less than 1-ms latency is also expected [12]. The most exciting feature of 6G is the inclusion of fully supported AI for driving autonomous systems. Video-type traffic is likely to be dominant among various data traffic systems in 6G communications. The most important technologies that will be the driving force for 6G are the terahertz (THz) band, AI, optical wireless communication (OWC), 3D networking, unmanned aerial vehicles (UAV), and wireless power transfer. In this paper, we describe how 6G communication systems can be developed; we also describe the expected 6G technologies, and the research issues required to address the needs of future smart networks.

The rest of the paper is organized as follows. Section II presents the growing trends in the use of mobile communications. The possible network architecture with the applications of future 6G communication systems is presented in Section III. The expected service requirements and the network characteristics in 6G communication systems are presented in Section IV. The possible key technologies for the development of 6G system are briefly discussed in Section V. In Section VI, we present the various research activities on 6G. We provide the key challenges and the research directions to reach the goal of 6G in Section VII. Finally, we draw our conclusions in Section VIII.

## II. TRENDS IN MOBILE COMMUNICATIONS

Since the beginning of the first analog communication system in the 1980s, a new generation of communication systems has been introduced almost every ten years. The transfer from one generation to another improves the QoS metrics, includes new services, and provides new features. During the last ten years, mobile data traffic has grown tremendously because of the introduction of smart devices and machine-to-machine (M2M) communications. Fig. 1 depicts the exponential growth of mobile connectivity. It is expected that the global mobile traffic volume will increase 670 times in 2030 as compared with the mobile traffic in 2010 [2]. The International Telecommunication Union (ITU) forecasted that by the end of 2030, the overall mobile data traffic volume will exceed 5 ZB per month. The number of mobile subscriptions will reach 17.1 billion as compared with 5.32 billion in 2010. Moreover, the use of M2M connectivity will also increase exponentially. The traffic volume for each of the mobile devices will also increase. The traffic volume of a mobile device in 2010 was 5.3 GB per month. However, this volume will increase 50 times in 2030. The number of M2M subscription will increase 33 times in 2020 and 455 times in 2030, as compared with 2010. Table 1 presents a few comparisons of the use of mobile connectivity in 2010, 2020, and 2030.

Recently, research interests have shifted to data-driven adaptive and intelligent methods. The upcoming 5G wireless networks will build a foundation of intelligent networks that provide AI operations [3]. It is estimated that by 2030, the capacity of 5G will reach its limit [12]. Then, fully intelligent network adaptation and management for providing advanced services will only be realized using 6G networks. Hence, 6G wireless communications is the result of the user needs growing beyond what the 5G network can offer. Researchers worldwide are already studying what 6G communications would be like in 2030; they are also looking at the possible drivers for successful 6G wireless communications. A few of the key motivating trends behind the evolution of 6G communication systems are as follows: high bit rate, high reliability, low latency, high energy efficiency, high spectral efficiency, new spectrums, green communication, intelligent networks, network availability, and convergence of communications, localization, computing, control, and sensing; 6G will be a fully digital and connected world.

**TABLE 1:** Global trends of wireless connectivity.

| Issue | 2010 | 2020 (Predicted) | 2030 (Predicted) | Unit |
|---|---|---|---|---|
| Mobile subscriptions | 5.32 | 10.7 | 17.1 | Billion |
| Smartphone subscriptions | 0.645 | 1.3 | 5.0 | Billion |
| M2M subscriptions | 0.213 | 7.0 | 97 | Billion |
| Traffic volume | 7.462 | 62 | 5016 | EB/month |
| M2M traffic volume | 0.256 | 5 | 622 | EB/month |
| Traffic per subscriber | 1.35 | 10.3 | 257.1 | GB/month |

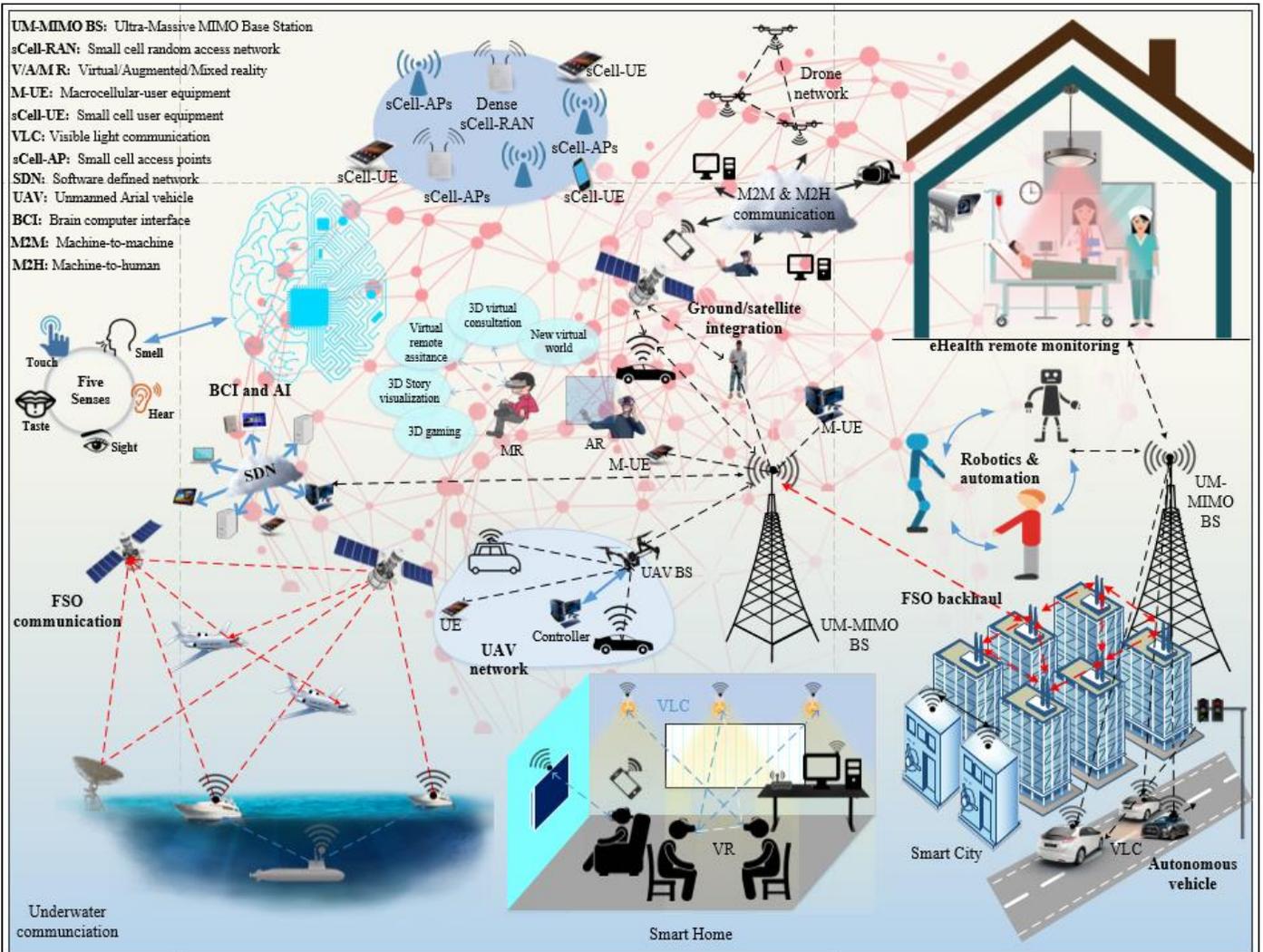

Fig. 2. Possible 6G communication architecture scenario.

### III. PROSPECTS AND APPLICATIONS

AI will be integrated into the 6G communication systems. All the network instrumentation, management, physical layer signal processing, resource management, service-based communications, and so on will be incorporated by using AI [12]. It will foster the Industry 4.0 revolution, which is the digital transformation of industrial manufacturing [13]. Fig. 2 shows the communication architecture scenario toward envisioning the 6G communication systems. Some key prospects and applications of 6G wireless communication are briefly described below.

*Super-smart society*: The superior features of 6G will accelerate the building of smart societies leading to life quality improvements, environmental monitoring, and automation using AI-based M2M communication and energy harvesting [13]. The 6G wireless connectivity will make our society super smart through the use of smart mobile devices, autonomous vehicles, and so on. In addition, many cities in the world will deploy flying taxis based on 6G wireless technology. Smart homes will become a reality because any device in a remote location will be controlled by using a command given from a smart device.

*Extended* reality: Extended reality (hereinafter XR) services including augmented reality (AR), mixed reality (MR), and VR are very important features of 6G communication systems. All these features use 3D objects and AI as their key driving elements. Besides providing perceptual requirements of computing, cognition, storage, human senses, and physiology, 6G will provide a truly immersive AR/MR/VR experience by joint design integration and high-quality 6G wireless connectivity [9]. VR is a computer-simulated 3D experience in which computer technologies use reality headsets to generate realistic sensations and replicate a real environment or create an imaginary world. A true VR environment engages all five senses. AR is a live view of a physical real world whose elements are augmented by various computer-generated sensor inputs, such as audio, video, visuals, and global positioning system (GPS) data. It uses the existing reality and adds to it by using a device of some sort. MR merges the real and the virtual worlds to create new atmospheres and visualizations to interact in real-time. It is also sometimes named as hybrid reality. One important characteristic of MR is that the artificial and real world contents can respond to one another in real-time. XR refers to all combined real and virtual environments and human–machine interactions generated by computer technology and wearables. It includes all its descriptive forms such as AR, VR, and MR. It brings together AR, VR, and MR under one term. The high-data-rate, low latency, and highly reliable wireless connectivity provided in the 6G system is very important for a genuine XR (i.e., AR, VR, and MR) experience.

*Connected robotics and autonomous systems*: Currently, a number of automotive technology researchers have been investigating automated and connected vehicles. The 6G systems help in the deployment of connected robots and autonomous systems. The drone-delivery UAV system is an example of such a system. The automated vehicle based on 6G wireless communication can dramatically change our daily lifestyles. The 6G system will promote the real deployment of self-driving cars (autonomous cars or driverless cars). A self-driving car perceives its surroundings by combining a variety of sensors, such as light detection and ranging (LiDAR), radar, GPS, sonar, odometry, and inertial measurement units. The 6G system will support reliable vehicle-to-everything and vehicle-to-server connectivity. A UAV is a type of unmanned aircraft. The ground-based controller and the system communications between the UAV and the ground will be supported by 6G networks. UAVs help in many fields such as military, commerce, science, agriculture, recreation, law and order, product delivery, surveillance, aerial photography, disaster management, and drone racing. Moreover, the UAV will be used to support wireless broadcast and high rate transmissions when the cellular base station (BS) is absent or not working [14].

*Wireless brain–computer interactions*: The brain–computer interface (BCI) is an approach to control the appliances that are used daily in smart societies, especially the appliances used at home and in medical systems [15], [16]. It is a direct communication path between the brain and external devices. BCI acquires the brain signals that transmit to a digital device and analyze and interpret the signals into further commands or actions. The features of 6G wireless communication will support the true implementation of BCI systems for living a smart life.

*Haptic Communication*: Haptic communication is a branch of nonverbal communication that uses the sense of touch. The proposed 6G wireless communication will support haptic communication; remote users will be able to enjoy haptic experiences through real-time interactive systems [17]. The implementation of haptic systems and applications will be facilitated by the superior features of 6G communication networks.

*Smart healthcare*: Medical health systems will also benefit by the 6G wireless systems because innovations, such as AR/VR, holographic telepresence, mobile edge computing, and AI, will help build smart healthcare systems [13]. A reliable remote monitoring system in the healthcare system will be facilitated by the 6G systems. Even remote surgery will be made possible by using 6G communication. A high-data-rate, low latency, and ultra-reliable 6G network will help to quickly and reliably transport huge volumes of medical data, which can improve both the access to care and the quality of care.

*Automation and manufacturing*: Full automation based on AI will be provided by 6G. The term "automation" refers to the automatic control of processes, devices, and systems. The 6G automation systems will provide highly reliable, scalable, and secure communications using high-data-rate and low latency networks. The 6G system will also provide network integrity because it ensures error-free data transfer without any data loss between transmission and reception.

*Five sense information transfer*: To experience the world around them, humans use their five senses of hearing, sight, taste, smell, and touch. The 6G communication systems will remotely transfer data obtained from the five senses. This technology uses the neurological process through sensory integration. It detects the sensations from the human body and the environment and uses the body effectively within the environment and local circumstances. The BCI technology will effectively boost this application.

*Internet of everything*: IoE is the seamless integration and autonomous coordination among a very large number of computing elements and sensors, objects or devices, people, processes, and data using the internet infrastructure [18]. The 6G system will provide full IoE support. It is basically a kind of internet of things (IoT), but it is an umbrella term that integrates the four properties, such as data, people, processes, and physical devices, in one frame [19]. IoT is generally about the physical devices or objects and communicating with one another, but IoE introduces network intelligence to bind all people, data, processes, and physical objects into one system. IoE will be used for smart societies, such as smart cars, smart health, and smart industries.

## IV. SPECIFICATIONS AND REQUIREMENTS

5G technologies have been associated with trade-offs of several issues such as throughput, delay, energy efficiency, deployment costs, reliability, and hardware complexity. It is

**Draft**

TABLE 2: Comparison of 6G with 4G and 5G communication systems

| Issue | 4G | 5G | 6G |
|---|---|---|---|
| Per device peak data rate | 1 Gbps | 10 Gbps | 1 Tbps |
| E2E latency | 100 ms | 10 ms | 1 ms |
| Maximum spectral efficiency | 15 bps/Hz | 30 bps/Hz | 100 bps/Hz |
| Mobility support | Up to 350 km/hr | Up to 500 km/hr | Up to 1000 km/hr |
| Satellite integration | No | No | Fully |
| AI | No | Partial | Fully |
| Autonomous vehicle | No | Partial | Fully |
| XR | No | Partial | Fully |
| Haptic Communication | No | Partial | Fully |

very likely that 5G will not be able to meet the market demands after 2030. Then, 6G will fill the gap between 5G and the market demand. Based on the previous trends and predictions of future needs, the main objectives for the 6G systems are (i) extremely high data rates per device, (ii) a very large number of connected devices, (iii) global connectivity, (iv) very low latency, (v) lowering the energy consumption with battery-free IoT devices, (vi) ultra-high reliable connectivity, and (vii) connected intelligence with machine learning capability. Table 2 shows a comparison of 6G with the 4G and 5G communication systems.

*A. Service Requirements*

The 6G wireless system will have the following key factors:
- Enhanced mobile broadband (eMBB)
- Ultra-reliable low latency communications (URLLC)
- Massive machine-type communication (mMTC)
- AI integrated communication
- Tactile internet
- High throughput
- High network capacity
- High energy efficiency
- Low backhaul and access network congestion
- Enhanced data security

It is estimated that the 6G system will have 1000 times higher simultaneous wireless connectivity than the 5G system. URLLC, which is a key 5G feature, will be a key driver again in 6G communication by providing end-to-end (E2E) delay of less than 1 ms [12]. Volume spectral efficiency, as opposed to the often-used area spectral efficiency, will be much better in 6G [12]. The 6G system will provide ultra-long battery life and advanced battery technology for energy harvesting. In 6G systems, mobile devices will not need to be separately charged.

*B. New Network Characteristics*

*Satellites integrated network*: To provide global mobile collectivity, 6G is expected to integrate with satellites. Integrating terrestrial, satellite, and airborne networks into a single wireless system will be crucial for 6G.

*Connected intelligence*: In contrast to the earlier generation of wireless communication systems, 6G will be transformative, and will update the wireless advancement from "connected things" to "connected intelligence" [20]. AI will be introduced in each step of the communication process. The pervasive introduction of AI will produce a new paradigm of communication systems.

*Seamless integration of wireless information and energy transfer*: The 6G wireless networks will also transfer power to charge battery devices, such as smartphones and sensors. Hence, wireless information and energy transfer (WIET) will be integrated.

*Ubiquitous super 3D connectivity*: Accessing the network and core network functionalities on drones and very low earth orbit satellites will make the super-3D connectivity in 6G ubiquitous.

*C. Few General Requirements in Network Characteristics*

*Small cell networks*: The idea of small cell networks has been introduced to improve the received signal quality as a consequence of throughput, energy efficiency, and spectral efficiency enhancement in cellular systems [21]–[23]. As a result, small cell networks are an essential characteristic for the 5G and beyond (5GB) communication systems. Therefore, 6G communication systems also adopt this network characteristic.

*Ultra-dense heterogeneous networks*: Ultra-dense heterogeneous networks [24], [25] will be another important characteristic of 6G communication systems. Multi-tier networks consisting of heterogeneous networks will improve the overall QoS and reduce the cost.

*High-capacity backhaul*: The backhaul connectivity must be characterized by high-capacity backhaul networks to support a huge volume of traffic. High-speed optical fiber and free space optical (FSO) systems are possible solutions for this problem.

*Radar technology integrated with mobile technologies*: High-accuracy localization with communication is also one of the features of the 6G wireless communication system. Hence, radar systems will be integrated with 6G networks.

*Softwarization and virtualization*: Softwarization and virtualization are two important features that are the basis of the design process in 5GB networks to ensure flexibility, reconfigurability, and programmability. Moreover, they will allow to share the billions of devices in a shared physical infrastructure.

V. KEY ENABLING TECHNOLOGIES OF 6G

The 6G system will be driven by many technologies. A few expected key technologies for 6G are discussed below.

*Artificial Intelligence*: The most important and newly introduced technology for 6G communication systems is AI [26]–[30]. There was no involvement of AI for 4G communication systems. The upcoming 5G will support partial or very limited AI. However, 6G will be fully supported by AI for automatization. Advancements in machine learning will create more intelligent networks for real-time communications in 6G. The introduction of AI in communication will simplify and improve the transport of real-time data. Using numerous analytics, AI can determine the way a complex target job is performed. AI will increase the efficiency and reduce the processing delay of the communication steps. Time-consuming tasks, such as handover and network selection, can be performed promptly by using AI. AI will also play a vital role in M2M, machine-to-human, and human-to-machine communications. It will also prompt communication in the BCI. AI-based communication systems will be supported by metamaterials, intelligent structures, intelligent networks, intelligent devices, intelligent cognitive radio, self-sustaining wireless networks, and machine learning.

*Terahertz communications*: Spectral efficiency can be increased by increasing the bandwidth; this can be done by using sub-THz communication with wide bandwidths and by applying advanced massive multiple input, multiple output (MIMO) technologies. The RF band has been almost exhausted, and now it is insufficient to meet the high demands of 6G. The THz band will play an important role in 6G communication [31], [32]. The THz band is intended to be the next frontier of high-data-rate communications. THz waves, also known as submillimeter radiation, usually refer to the frequency band between 0.1 THz, and 10 THz with the corresponding wavelengths in the 0.03 mm–3 mm range [33]. According to the recommendations of ITU-R, the 275 GHz–3 THz band range is considered to be the main part of the THz band for cellular communications [33]. The capacity of 6G cellular communications will be increased by adding the THz band (275 GHz–3THz) to the mmWave band (30–300 GHz). The band within the range of 275 GHz–3 THz has not yet been allocated for any purpose worldwide; therefore, this band has the potential to accomplish the desired high data rates [32]. When this THz band is added to the existing mmWave band, the total band capacity increases a minimum of 11.11 times. Of the defined THz bands, 275 GHz–3THz, and 275 GHz–300 GHz lie on the mmWave, and 300 GHz–3 THz lie on the far infrared (IR) frequency band. Even though the 300 GHz–3 THz band is part of the optical band, it is at the boundary of the optical band and immediately after the RF band. Hence, this 300 GHz–3 THz band shows similarities with the RF. THz heightens the potentials and challenges of high frequency communications [4].

The key properties of THz communications include (i) widely available bandwidth to support very high data rates (ii) high path loss arising from the high frequency (highly directional antennas will most probably be indispensable) [1]. The narrow beamwidths generated by the highly directional antennas reduce the interference. The small wavelength of the THz signals allows a much greater number of antenna elements to be incorporated into the devices and BSs operating in this band. This allows the use of advanced adaptive array technologies that can overcome the range limitations [1].

*Optical wireless technology:* OWC technologies are envisioned for 6G communications in addition to RF-based communications for all possible device-to-access networks; these networks also access network-to-backhaul/fronthaul network connectivity. OWC technologies are already being used since 4G communication systems. However, it will be used more widely to meet the demands of 6G communication systems. OWC technologies, such as light fidelity, visible light communication, optical camera communication, and FSO communication based on the optical band are already well-known technologies [34]–[37]. Researchers have been working on enhancing the performance and overcoming the challenges of these technologies. Communication based on optical wireless technologies can provide very high data rates, low latencies, and secure communications. LiDAR, which is also based on the optical band, is a promising technology for very-high-resolution 3D mapping in 6G communications.

*FSO backhaul network*: It is not always possible to have optical fiber connectivity as a backhaul network because of remote geographical locations and complexities. The FSO backhaul network is very promising for 5GB communication systems [38]–[40]. The transmitter and receiver characteristics of the FSO system are similar to those of optical fiber networks. Therefore, the data transfer in the FSO system is comparable with the optical fiber system. Hence, along with the optical fiber networks, FSO is an excellent technology for providing backhaul connectivity in 6G. Using FSO, it is possible to have very long range communications even at a distance of more than 10,000 km. FSO supports high-capacity backhaul connectivity for remote and non-remote areas, such as the sea, outer space, underwater, isolated islands; FSO also supports cellular BS connectivity.

*Massive multiple input, multiple output technique*: One key technique to improve spectral efficiency is the application of the MIMO technique [41], [42]. When the MIMO technique improves, the spectral efficiency also improves. Therefore, a massive MIMO technology will be crucial in the 6G system.

*Blockchain:* Blockchain will be an important technology to manage massive data in future communication systems [43]–[45]. Blockchains are just one form of the distributed ledger technology. A distributed ledger is a database that is distributed across numerous nodes or computing devices. Each node replicates and saves an identical copy of the ledger. The blockchain is managed by peer-to-peer networks. It can exist without being managed by a centralized authority or a server. The data on a blockchain is gathered together and structured in blocks. The blocks are connected to one another and secured using cryptography. The blockchain is essentially a perfect complement to the massive IoT with improved interoperability, security, privacy, reliability, and scalability [46]. Therefore, the blockchain technology will provide several facilities, such as interoperability across devices, traceability of massive data, autonomic interactions of different IoT systems, and reliability for the massive connectivity of 6G communication systems.

*3D networking*: The 6G system will integrate the ground and airborne networks to support communications for users in the vertical extension. The 3D BSs will be provided through low orbit satellites and UAVs [47], [48]. The addition of new dimensions in terms of altitude and related degrees of freedom makes 3D connectivity considerably different from the conventional 2D networks.

*Quantum communications:* Unsupervised reinforcement learning in networks is promising in the context of 6G networks. Supervised learning approaches will not feasible for labeling huge volumes of data generated in 6G. Unsupervised learning does not need labeling. Hence, this technique can be used for autonomously building the representations of complex networks. By combining reinforcement learning and unsupervised learning, it is possible to operate the network in a truly autonomous fashion [3].

*Unmanned aerial vehicle:* UAVs or drones will be an important element in 6G wireless communications. In many cases, high-data-rate wireless connectivity will be provided using the UAV technology. The BS entities will be installed in UAVs to provide cellular connectivity. A UAV has certain features that are not found in fixed BS infrastructures, such as easy deployment, strong line-of-sight links, and degrees of freedom with controlled mobility [12]. During emergency situations, such as natural disasters, the deployment of terrestrial communication infrastructures is not economically feasible, and sometimes it is not possible to provide any service in volatile environments. UAVs can easily handle these situations. A UAV will be the new paradigm in the field of wireless communication. This technology can facilitate three fundamental requirements of wireless networks that are, eMBB, URLLC, and mMTC [14]. UAVs can also serve several purposes, such as the enhancement of network connectivity, fire detection, emergency services in disaster, security and surveillance, pollution monitoring, parking monitoring, accident monitoring, and so on. Therefore, UAV technology is recognized as one of the most important technologies for 6G communication.

*Cell-free communications*: The tight integration of multiple frequencies and heterogeneous communication technologies will be crucial in 6G systems. As a result, the user will move seamlessly from one network to another network without the need for making any manual configurations in the device [4]. The best network will be automatically selected from the available communication technology. This will break the limits of the concept of cells in wireless communications. Currently, the user movement from one cell to another cell causes too many handovers in dense networks, and also causes handover failures, handover delays, data losses, and the ping-pong effect. The 6G cell-free communications will overcome all these and provide better QoS. Cell-free communication will be achieved through multi-connectivity and muti-tier hybrid techniques and by different and heterogeneous radios in the devices [4].

*Integration of wireless information and energy transfer:* WIET in communication will be one of the most innovative technologies in 6G. WIET uses the same fields and waves as wireless communication systems. In particular, sensors and smartphones will be charged by using wireless power transfer during communication. WIET is a promising technology for lengthening the lifetime of the battery-charging wireless systems [49]. Hence, devices without batteries will be supported in 6G communications.

*Integration of sensing and communication*: A key driver for autonomous wireless networks is the capability to continuously sense the dynamically changing states of the environment and exchange the information among different nodes [50]. In 6G, the sensing will be tightly integrated with communication to support autonomous systems.

*Integration of access-backhaul networks*: The density of the access networks in 6G will be huge. Each access network is connected with backhaul connectivity, such as optical fibers and FSO networks. To cope with the very large number of access networks, there will be tight integration between the access and backhaul networks.

*Dynamic network slicing*: Dynamic network slicing permits a network operator to allow dedicated virtual networks to support the optimized delivery of any service toward a wide range of users, vehicles, machines, and industries. It is one of the most important elements for management when a large number of users are connected to a large number of heterogeneous networks in 5GB communication systems.

*Holographic beamforming*: Beamforming is a signal processing procedure by which an array of antennas can be steered to transmit radio signals in a specific direction. It is a subset of smart antennas or advanced antenna systems. The beamforming technique has several advantages, such as a high signal-to-noise ratio, interference prevention, and rejection, and high network efficiency. Holographic beamforming (HBF) is a new method for beamforming that is considerably different from the MIMO systems because it uses software-defined antennas. HBF will be a very effective approach in 6G for the efficient and flexible transmission and reception of signals in multi-antenna communication devices.

*Big data analytics*: Big data analytics is a complex process for analyzing a variety of large data sets or big data. This process uncovers information, such as hidden patterns, unknown correlations, and customer inclinations, to ensure perfect data management. The big data is collected from a wide variety of sources, such as videos, social networks, images, and sensors. This technology will be widely used for handling of huge data in 6G systems.

## VI. STANDARDIZATION AND RESEARCH ACTIVITIES

The 5G specifications have already been prepared, and even though it has already been launched in some parts of the world, the full phase of 5G will be deployed in 2020. Research activities on 6G are in their initial stages. From 2020, a number of studies will be performed worldwide on the standardization of 6G; 6G communication is still in its infancy. Many researchers have defined 6G as B5G or 5G+. Preliminary research activities have already started in the United States of America. The US president has requested the deployment of 6G in the country. China has already started the concept study for the development and standardization of 6G communications in 2019. The Chinese are planning for active research work on 6G in 2020. Most European countries, Japan, and Korea are planning several 6G projects. The research activities on 6G are expected to start in 2020. In this section, we present a few research activities, and standardization efforts. Table 3 presents the summary of a few studies on 6G communication.

TABLE 3: Summary of current studies on 6G

| Reference | Contribution and Research Direction |
|---|---|
| [3] | A comprehensive review of machine learning, quantum computing, and quantum machine learning and their potential benefits, issues, and use cases for their applications in the B5G networks is presented. A quantum computing assisted framework for 6G communication networks is also proposed. |
| [4] | Few 6G use cases were presented. It also tries to estimate 6G requirements. |
| [9] | The primary drivers of 6G applications and accompanying technological trends are discussed. In addition, a set of service classes and their target 6G performance requirements are proposed. |
| [12] | An extended vision of 5G from the viewpoint of the required changes for enabling 6G is presented. |
| [14] | A comprehensive survey on UAV communications toward 5G/B5G wireless networks is presented. |
| [20] | Possible technologies to enable mobile AI applications for 6G and AI-enabled approaches for 6G network design and optimization are discussed. |
| [26] | AI revolution for 6G wireless networks is discussed. |
| [27] | The role of Edge AI in future 6G communication is discussed. |
| [28] | The use of modern random access for IoT applications in 6G is proposed. In addition, a short overview of the recent advances in uncoordinated medium access is provided. |
| [29] | A vision for machine-type communication in 6G is discussed. Some relevant performance indicators and a few technologies are also presented. |
| [30] | A survey of intelligent reflecting surfaces for 6G wireless communications is presented. |
| [51] | The consequence of temporal correlation for joint success probability and the distribution of the number of interferers for UAV networks is analyzed. |
| [52] | The theoretical performance limits of reconfigurable intelligent surface-assisted communication systems on the potential use cases of intelligent surfaces in 6G wireless networks is presented. |
| [53] | A time-efficient neighbor discovery protocol for the THz band communication networks is proposed. |
| [54] | A new multiple-access method and delta-orthogonal multiple access for massive access in 6G cellular networks is proposed |
| [55] | The potential directions to achieve further flexibility in radio access technologies B5G is discussed. In addition, a framework for developing flexible waveform, numerology, and frame design strategies is also proposed. |
| [56] | A pair of bottlenecks that severely limit the integrated space and terrestrial network capacity are discussed. In addition, a review of the family of wireless communication technologies suitable for supporting such backbone links are presented. |
| [57] | The potential 6G techniques and an overview of the latest research on the promising techniques evolving for 6G are discussed. |
| [58] | Some technical challenges and opportunities for wireless communication and sensing applications above 100 GHz is presented. Moreover, a number of discoveries, approaches, and recent results that will help in the development and implementation of 6G wireless networks is also discussed. |
| [59] | A list of technical topics considering B5G is presented. |
| [60] | The benefit of data centers from 6G technologies, particularly in assuring flexibility and adaptability is presented. |

*Samsung Electronics*: Samsung Electronics has opened an R&D center for the development of essential technologies for 6G mobile networks. To accelerate the development of solutions and for the standardization of 6G, Samsung research is conducting extensive research on cellular technologies; they have upgraded a next-generation telecommunication research team to a center.

*Finnish 6G Flagship program*: University of Oulu began the 6G research activities under Finland's flagship program. Research in 6G Flagship is organized into four unified planned research parts: wireless connectivity, distributed computing, services, and applications. Scientific innovations will be developed for important technology components of 6G systems.

*International Telecommunication Union*: Standardization activities on 5G of the ITU radio communication sector ITU-R was based on IMT-2020. Consequently, ITU-R will probably release IMT-2030, which will summarize the possible requirements of mobile communications in 2030 (i.e., 6G).

*6G wireless summit*: A successful first 6G wireless summit was held in Lapland, Finland, in March 2019. A wide and effective discussion was performed by academicians, industry persons, and vendors from all over the world. Pioneering wireless communication researchers were present at the summit. Moreover, the world's leading telecom companies also attended the summit. This 6G summit initiates the discussions on key issues, such as the motivation behind 6G, the way to move from 5G to 6G, the current industry trends for 6G, and the enabling technologies.

VII. CHALLENGES AND FUTURE RESEARCH DIRECTIONS

Several technical problems need to be solved to successfully deploy 6G communication systems. A few of possible concerns are briefly discussed below.

*High propagation and atmospheric absorption of THz*: The high THz frequencies provide high data rates. However, the THz bands need to overcome an important challenge for data transfer over relatively long distances because of the high propagation loss, and atmospheric absorption characteristics [1]. We require a new design for the transceiver architecture for the THz communication systems. The transceiver must be able to operate at high frequencies, and we need to ensure the full use of very widely available bandwidths. A very small gain and an effective area of the distinct THz band antennas is another challenge of THz communication. Health and safety concerns related to THz band communications also need to be addressed.

*Complexity in resource management for 3D networking*: The 3D networking extended in the vertical direction. Hence, a new dimension was added. Moreover, multiple adversaries may intercept legitimate information, which may significantly degrade the overall system performance. Therefore, new techniques for resource management and optimization for mobility support, routing protocol, and multiple access are essential. Scheduling needs a new network design.

*Heterogeneous hardware constraints*: In 6G, a very large number of heterogeneous types of communication systems, such as frequency bands, communication topologies, service delivery, and so on, will be involved. Moreover, the access points and mobile terminals will be significantly different in the hardware settings. The massive MIMO technique will be further upgraded from 5G to 6G, and this might require a more complex architecture. It will also complicate the communication protocol and the algorithm design. However, machine learning and AI will be included in communication. Moreover, the hardware design for different communication systems is different. Unsupervised and reinforcement learning may create complexities in hardware implementation as well. Consequently, it will be challenging to integrate all the communication systems into a single platform.

*Autonomous wireless systems*: The 6G system will provide full support to automation systems such as autonomous car, UAVs, and Industry 4.0 based on AI. To make autonomous wireless systems, we need to have the convergence of many heterogeneous sub-systems, such as autonomous computing, interoperable processes, system of systems, machine learning, autonomous cloud, machines of systems, and heterogeneous wireless systems [60]. Thus, the overall system development becomes complex and challenging. For example, developing a fully autonomous system for the driverless vehicle will be much more challenging because 6G researchers need to design fully automated self-driving vehicles that perform better than the human-controlled vehicles.

*Modeling of sub-mmWave (THz) frequencies*: The propagation characteristics of the mmWave and sub-mmWave (THz) is subject to atmospheric conditions; therefore, absorptive and dispersive effects are seen [61]. The atmospheric condition is frequently changeable and thus quite unpredictable. Therefore, the channel modeling of this band is relatively complex, and this band does not have any perfect channel model.

*Device capability*: The 6G system will provide a number of new features. Devices, such as smartphones, should have the capability to cope with the new features. In particular, it is challenging to support 1 Tbps throughput, AI, XR, and integrated sensing with communication features using individual devices. The 5G devices may not support few of the 6G features, and the capability improvement in 6G devices may increase the cost as well. There will be billions of devices connected to the 5G technology; therefore, we need to ensure that those devices are compatible with the 6G technology also.

*High-capacity backhaul connectivity*: The access networks in 6G will have a very high density. Moreover, these access networks are diverse in nature and widespread within a geographical location. Each of these access networks will support very high-data-rate connectivity for diverse types of users. The backhaul networks in 6G must handle the huge volume of data for connecting between the access networks and the core network to support high-data-rate services at the user level; otherwise, a bottleneck will be created. The optical fiber and FSO networks are possible solutions for high-capacity backhaul connectivity; therefore, any improvement in the capacity of these networks is challenging for the exponentially growing data demands of 6G.

*Spectrum and interference management:* Due to the scarcity of the spectrum resources and interference issues, it is very important to efficiently manage the 6G spectra including the spectrum-sharing strategies and innovative spectrum management techniques. Efficient spectrum management is important for achieving the maximum resource utilization with QoS maximization. In 6G, researchers have to address concerns, such as how to share the spectrum, and how to manage the spectrum mechanism in the heterogeneous networks that synchronize the transmission at the same frequency. Researchers also need to investigate how the interference can be cancelled using the standard interference cancellation methods, such as parallel interference cancellation, and successive interference cancellation.

*Beam management in THz communications*: Beamforming through massive MIMO systems is promising technology for supporting high-data-rate communications. However, beam management in sub-mmWave, that is, the THz band is challenging because of the propagation characteristics of the sub-mmWave. Hence, efficient beam management against unfavorable propagation characteristics will be challenging for massive MIMO systems [38]. However, for seamless handover, it is also important to choose the optimal beam efficiently in high-speed vehicular systems.

VIII. CONCLUSION

Each generation of communication system brings new and exciting features. The 5G communication system, which will be officially launched worldwide in 2020, has exciting features. However, 5G will not be able to fully support the growing demand for wireless communication in 2030. Therefore, 6G will need to be rolled out. Research on 6G is still in its infancy and in the study phase. This paper envisions the prospects and ways to reach the goal of 6G communication. In this paper, we presented the possible applications and the technologies to be deployed for 6G communication. We also described the possible challenges and research directions to reach the goals for 6G. Besides clarifying the vision and goal of 6G communications, we have stated the various technologies that could be used for 6G communication.

**References**

[1] S. Mumtaz *et al.*, "Teraherz communication for vehicular networks," *IEEE Transactions on Vehicular Technology*, vol. 66, no. 7, pp. 5617-5625, July 2017.
[2] ITU-R M.2370-0, *IMT traffic estimates for the years 2020 to 2030*, Jul. 2015.
[3] S. J. Nawaz, S. K. Sharma, S. Wyne, M. N. Patwary, and M. Asaduzzaman, "Quantum machine learning for 6G communication networks: state-of-the-art and vision for the future," *IEEE Access*, vol. 7, pp. 46317-46350, 2019.
[4] M. Giordani *et al*, "Towards 6G networks: use cases and technologies," *arXiv*:1903.12216
[5] M. Shafi et al., "5G: a tutorial overview of standards, trials, challenges, deployment, and practice," *IEEE Journal on Selected Areas in Communications,* vol. 35, no. 6, pp. 1201-1221, Jun. 2017.
[6] D. Zhang, Z. Zhou, S. Mumtaz, J. Rodriguez, and T. Sato, "One integrated energy efficiency proposal for 5G IoT communications," *IEEE Internet of Things Journal,* vol. 3, no. 6, pp. 1346-1354, Dec. 2016.
[7] M. Jaber, M. A. Imran, R. Tafazolli, and A. Tukmanov, "5G backhaul challenges and emerging research directions: a survey," *IEEE Access,* vol. 4, pp. 1743-1766, Apr. 2016.
[8] J. G. Andrews, S. Buzzi, W. Choi, S. V. Hanly, A.Lozano, A. C. K. Soong, and J. C. Zhang, "What will 5G be?," *IEEE Journal on Selected Areas in Communications,* vol. 32, no. 6, pp. 1065-1082, Jun. 2014.
[9] W. Saad, M. Bennis, and M. Chen, "A vision of 6G wireless systems: applications, trends, technologies, and open research problems," *arXiv*:1902.10265
[10] 123 Seminars Only. (2019). *6G mobile technology*. [Online]. Available: http://www.123seminarsonly.com/CS/6G-Mobile-Technology.html
[11] K. David and H. Berndt, "6G vision and requirements: is there any need for beyond 5G?," *IEEE Vehicular Technology Magazine*, vol. 13, no. 3, pp. 72-80, Sept. 2018.
[12] F. Tariq *et al*., "A Speculative Study on 6G," *arXiv*:1902.06700.
[13] (2019). *6G*. [Online]. Available: http://mmwave.dei.unipd.it/research/6g/
[14] B. Li, Z. Fei, and Y. Zhang, "UAV communications for 5G and beyond: recent advances and future trends," *IEEE Internet of Things Journal*, vol. 6, no. 2, pp. 2241-2263, April 2019.
[15] A. K. Tripathy, S. Chinara, and M. Sarkar, "An application of wireless brain–computer interface for drowsiness detection," *Biocybernetics and Biomedical Engineering*, vol. 36, no. 1, pp. 276-284, 2016.
[16] S. R. A. Jafri, *et al*, "Wireless brain computer interface for smart home and medical system," *Wireless Personal Communications*, vol. 106, no. 4, pp. 2163-2177, Jun. 2019.
[17] PR Newswire. (2019). *5G and the haptic internet: emerging technologies, solutions and market opportunities*. [Online]. Available: https://www.prnewswire.com/news-releases/5g-and-the-haptic-internet-emerging-technologies-solutions-and-market-opportunities-300184874.html
[18] Internet of Everything. (2019). *Internet of everything (IoE)*. [Online]. Available: https://ioe.org/
[19] CISCO. (2019). *The internet of everything*. [Online]. Available: https://www.cisco.com/c/dam/en_us/about/business-insights/docs/ioe-value-at-stake-public-sector-analysis-faq.pdf
[20] K. B. Letaief *et al*, "The roadmap to 6G - AI empowered wireless networks," *arXiv:1904.11686*
[21] M. Z. Chowdhury, M. T. Hossan, and Y. M. Jang, "Interference management based on RT/nRT traffic classification for FFR-aided small cell/macrocell heterogeneous networks," *IEEE Access,* vol. 6, pp. 31340-31358, Jun. 2018.
[22] A.S.M. Zadid Shifat, M, Z. Chowdhury, and Y. M. Jang, "Game-based approach for QoS provisioning and interference management in heterogeneous networks," *IEEE Access,* vol. 6, pp. 10208–10220, Jan. 2018.
[23] A. J. Mahbas, H. Zhu, and J. Wang, "Impact of small cells overlapping on mobility management," *IEEE Transactions on Wireless Communications*, vol. 18, no. 2, pp. 1054-1068, Feb. 2019.
[24] T. Zhou, N. Jiang, Z. Liu, and C. Li, "Joint cell activation and selection for green communications in ultra-dense heterogeneous networks," *IEEE Access*, vol. 6, pp. 1894-1904, 2018.
[25] S. Andreev, V. Petrov, M. Dohler, and H. Yanikomeroglu, "Future of ultra-dense networks beyond 5G: harnessing heterogeneous moving cells," *IEEE Communications Magazine*, vol. 57, no. 6, pp. 86-92, Jun. 2019.
[26] R.-A. Stoica and G. T. F. Abreu, "6G: the wireless communications network for collaborative and AI applications," *arXiv*:1904.03413.
[27] L. Loven *et al*, "Edge AI: a vision for distributed, edge-native artificial intelligence in future 6G networks," *6G Wireless Summit*, Levi, Finland, Mar. 2019.
[28] F. Clazzer *et al*, "From 5G to 6G: has the time for modern random access come?," *arXiv*:1903.03063
[29] N. H. Mahmood *et al*, "Six key enablers for machine type communication in 6G," *arXiv:1903.05406*
[30] J. Zhao, "a survey of reconfigurable intelligent surfaces: towards 6G wireless communication networks with massive MIMO 2.0," *arXiv:1907.04789*
[31] I. F. Akyildiz, J. M. Jornet, and C. Han, "Teraherz band: next frontier for wireless communications," *Physical Communication*, vol. 12, Sep. 2014, pp. 16-32.
[32] K. Tekbıyık, A. R. Ekti, G. K. Kurt, and A. Görçinad, "Teraherz band communication systems: challenges, novelties and standardization efforts," Physical Communication, vol. 35, Aug. 2019.
[33] ITU-R SM.2352-0, *Technology trends of active services in the frequency range 275-3 000 GHz*, Jun. 2015.
[34] M. Z. Chowdhury, M. T. Hossan, A. Islam, and Y. Min Jang, "A comparative survey of optical wireless technologies: architectures and applications," *IEEE Access,* vol. 6, pp. 9819–10220, Jan. 2018.
[35] M. Z. Chowdhury, M. T. Hossan, M. K. Hasan, and Y. M. Jang, "Integrated RF/optical wireless networks for improving QoS in indoor and transportation applications," *Wireless Personal Communications*, vol. 107, no. 3, pp. 1401-1430, Aug. 2019.
[36] M. T. Hossan *et al.*, "A new vehicle localization scheme based on combined optical camera communication and photogrammetry," *Mobile Information Systems,* vol. 2018, March 2018.
[37] M. T. Hossan, M. Z. Chowdhury, M. Shahjalal, and Y. M. Jang, "Human bond communication with head-mounted displays: scope, challenges, solutions, and applications," *IEEE Communications Magazine*, vol. 57, no. 2, pp. 26-32, February 2019.
[38] Z. Gu, J. Zhang, Y. Ji, L. Bai, and X. Sun, "Network topology reconfiguration for FSO-based fronthaul/backhaul in 5G+ wireless networks," *IEEE Access*, vol. 6, pp. 69426-69437, 2018.
[39] A. Douik, H. Dahrouj, T. Y. Al-Naffouri, and M. Alouini, "Hybrid radio/free-space optical design for next generation backhaul systems," *IEEE Transactions on Communications*, vol. 64, no. 6, pp. 2563-2577, Jun. 2016.
[40] B. Bag, A. Das, I. S. Ansari, A. Prokeš, C. Bose, and A. Chandra, "Performance analysis of hybrid FSO systems using FSO/RF-FSO link

**Draft**

[40] ...adaptation," *IEEE Photonics Journal*, vol. 10, no. 3, pp. 1-17, Jun. 2018, Art no. 7904417.

[41] H. Gao, Y. Su, S. Zhang, and M. Diao, "Antenna selection and power allocation design for 5G massive MIMO uplink networks," *China Communications*, vol. 16, no. 4, pp. 1-15, April 2019.

[42] M. Attarifar, A. Abbasfar, and A. Lozano, "Modified conjugate beamforming for cell-free massive MIMO," *IEEE Wireless Communications Letters*, vol. 8, no. 2, pp. 616-619, April 2019.

[43] R. Henry, A. Herzberg, and A. Kate, "Blockchain access privacy: challenges and directions," *IEEE Security & Privacy*, vol. 16, no. 4, pp. 38-45, Jul./Aug. 2018.

[44] T. Aste, P. Tasca, and T. Di Matteo, "Blockchain technologies: the foreseeable impact on society and industry," *Computer*, vol. 50, no. 9, pp. 18-28, 2017.

[45] D. Miller, "Blockchain and the internet of things in the industrial sector," *IT Professional*, vol. 20, no. 3, pp. 15-18, May/Jun. 2018.

[46] H.-N. Dai, Z. Zheng, and Y. Zhang, "Blockchain for internet of things: a survey," *arXiv:1906.00245*.

[47] C. Pan, J. Yi, C. Yin, J. Yu and X. Li, "Joint 3D UAV placement and resource allocation in software-defined cellular networks with wireless backhaul," *IEEE Access*, vol. 7, pp. 104279-104293, 2019.

[48] M. Mozaffari, A. Taleb Zadeh Kasgari, W. Saad, M. Bennis, and M. Debbah, "Beyond 5G with UAVs: foundations of a 3D wireless cellular network," *IEEE Transactions on Wireless Communications*, vol. 18, no. 1, pp. 357-372, Jan. 2019.

[49] H. Wang, W. Wang, X. Chen, and Z. Zhang, "Wireless information and energy transfer in interference aware massive MIMO systems," in *Proc. IEEE Global Communications Conference*, Austin, TX, 2014, pp. 2556-2561.

[50] M. Kobayashi, G. Caire, and G. Kramer, "Joint state sensing and communication: optimal tradeoff for a memoryless case," *arXiv:1805.05713*.

[51] M. Salehi and E. Hossain, "On the effect of temporal correlation on joint success probability and distribution of number of interferers in mobile UAV networks," *IEEE Wireless Communications Letters*.

[52] E. Basar, M. Di Renzo, J. de Rosny, M. Debbah, M. Alouini, and R. Zhang, "Wireless communications through reconfigurable intelligent surfaces," *IEEE Access*, Aug. 2019.

[53] Q. Xia and J. M. Jornet, "Expedited neighbor discovery in directional terahertz communication networks enhanced by antenna side-lobe information," *IEEE Transactions on Vehicular Technology*, vol. 68, no. 8, pp. 7804-7814, Aug. 2019.

[54] Y. Al-Eryani and E. Hossain, "The D-OMA method for massive multiple access in 6G: performance, security, and challenges," *IEEE Vehicular Technology Magazine*, Jul. 2019.

[55] Z. E. Ankarali, B. Peköz, and H. Arslan, "Flexible radio access beyond 5G: a future projection on waveform, numerology, and frame design principles," *IEEE Access*, vol. 5, pp. 18295-18309, 2017.

[56] X. Huang, J. A. Zhang, R. P. Liu, Y. J. Guo, and L. Hanzo, "Airplane-aided integrated networking for 6G wireless: will it work?," in *IEEE Vehicular Technology Magazine*, Jul. 2019.

[57] P. Yang, Y. Xiao, M. Xiao, and S. Li, "6G wireless communications: vision and potential techniques," *IEEE Network*, vol. 33, no. 4, pp. 70-75, July/August 2019.

[58] T. S. Rappaport *et al.*, "Wireless communications and applications above 100 GHz: opportunities and challenges for 6G and beyond," *IEEE Access*, vol. 7, pp. 78729-78757, 2019.

[59] M. Katz, M. Matinmikko-Blue, and M. Latva-Aho, "6Genesis flagship program: building the bridges towards 6G-enabled wireless smart society and ecosystem," in *proc. Latin-American Conference on Communications (LATINCOM)*, Guadalajara, 2018, pp. 1-9.

[60] D. Elliott, W. Keen, and L. Miao, "Recent advances in connected and automated vehicles," *Journal of Traffic and Transportation Engineering*, vol. 6, no. 2, pp. 109-131, Apr. 2019.

[61] Y. Golovachev, A. Etinger, G. A. Pinhasi, and Y. Pinhasi, "Propagation properties of sub-millimeter waves in foggy conditions," *Journal of Applied Physics*, Apr. 2019.

**Mostafa Zaman Chowdhury** obtained his B.Sc. degree in Electrical and Electronic Engineering from Khulna University of Engineering and Technology (KUET), Bangladesh, in 2002. He obtained his M.Sc. and Ph.D. degrees both in Electronics Engineering from Kookmin University, Korea, in 2008 and 2012, respectively. In 2003, he joined the Electrical and Electronic Engineering department at KUET as a Lecturer where he is currently affiliated as Professor. He has been working as a postdoc researcher at Kookmin University, Korea since 2017. He is a senior member of IEEE. In 2008, he received the Excellent Student Award from Kookmin University. He has published around 125 research papers in national and international conferences and journals. His three papers received the Best Paper Award at several international conferences around the world. He received the Best Reviewer Award 2018 by ICT Expressed journal. Moreover, he received the Education and Research Award 2018 given by Bangladesh Community in Korea. He served as a reviewer for many international journals (including IEEE, Elsevier, Springer, ScienceDirect, and Hindawi published journals) and IEEE conferences. He has been working as an Editor for ICT Express, Associate Editor of IEEE Access, Lead Guest Editor for Wireless Communications and Mobile Computing, and Guest Editor for Applied Sciences. He was a TPC chair of the International Workshop on 5G/6G Mobile Communications in 2017 and 2018. He was the publicity chair of the International Conference on Artificial Intelligence in Information and Communication, 2019. He has served as a TPC member for several IEEE conferences. He has been involved in several Korean government projects. His research interests include convergence networks, QoS provisioning, small-cell networks, Internet of Things, eHealth, 5G and beyond (5GB) communications, and optical wireless communication.

**Md. Shahjalal** obtained his B.Sc. degree in Electrical and Electronic Engineering from Khulna University of Engineering and Technology, Bangladesh, in 2017. He obtained his M.Sc. degree in Electronics Engineering from Kookmin University, Korea, in 2019. He is currently pursuing his Ph.D. degree in the Department of Electronics Engineering, Kookmin University, South Korea. His research interests include optical wireless communications, optical camera communication, NOMA, software-defined networking, deep neural network, and 5G.

**Shakil Ahmed** obtained his BS degree in Electrical and Electronic Engineering from Khulna University of Engineering and Technology (KUET), Bangladesh, in 2014. He received his MS degree in Electrical Engineering from Utah State University, Logan, Utah, the USA in 2019. Currently, he is pursuing his Ph.D. in Electrical Engineering at the University of Arizona, Tucson, Arizona, USA. He is a student member of IEEE. He received the prestigious Presidential Doctoral Research Fellowship by the school of graduate studies at Utah State University. He has published multiple research papers in international conferences and journals. One of his papers received the Best Paper Award at the international conference. He served as a reviewer for international journals, such as IEEE Access, Wireless Communications, and Mobile Computing, etc. His current research interests include next-generation wireless communications, wireless network design and optimization, unmanned aerial vehicle (UAV), physical layer security, and covert/low probability detection (LDP).

**Yeong Min Jang** received his B.E. and M.E. degrees in Electronics Engineering from Kyungpook National University, South Korea, in 1985 and 1987, respectively, and his Doctoral degree in computer science from the University of Massachusetts, USA, in 1999. He was with the Electronics and Telecommunications Research Institute (ETRI) from 1987 to 2000. Since 2002, he has been with the School of Electrical Engineering, Kookmin University, Seoul, South Korea, where he has been the Director of the Ubiquitous IT Convergence Center between 2005 and 2010, the Director of the LED Convergence Research Center since 2010, and the Director of the Internet of Energy Research Center since 2018. He is currently a life member of the Korean Institute of Communications and Information Sciences (KICS).

**Draft**

He received the Young Science Award from the Korean Government (2003–2006). He has organized several conferences and workshops, such as the International Conference on Ubiquitous and Future Networks (2009–2017), the International Conference on ICT Convergence (2010–2016), the International Conference on Information Networking 2015, and the International Workshop on Optical Wireless LED Communication Networks (2013–2016). He had served as the Founding Chair of the KICS Technical Committee on Communication Networks in 2007 and 2008. He had served as the Executive Director of KICS from 2006 to 2014, Vice President of KICS from 2014 to 2016, and Executive Vice President of KICS for 2018. He is the president of KICS for 2019. He serves as the Co-Editor-in-Chief of ICT Express, which is published by Elsevier. He had been the Steering Chair of the Multi-Screen Service Forum since 2011 and the Society Safety System Forum since 2015. He had served as the Chairman of the IEEE 802.15 Optical Camera Communications Study Group in 2014 and also served as the Chairman of the IEEE 802.15.7m Optical Wireless Communications Task Group. He is currently the Chairman of IEEE 802.15 Vehicular Assistive Technology (VAT) Interest Group. His research interests include 5G/6G mobile communications, Internet of Energy, eHealth, multiscreen convergence, public safety, optical wireless communications, optical camera communication, and the Internet of Things.